# Veiled Nonlocality and Cosmic Censorship

Menas Kafatos[1,a] and Subhash Kak[2]
1. Chapman University, Orange, CA 92866, USA
2. Oklahoma State University, Stillwater, OK 74078, USA

**Abstract.** The premise that consciousness has a quantum mechanical basis or correlate implies that its workings have a nonlocal component. To check whether consciousness as an entity leaves a physical trace, we propose that laboratory searches for such a trace should be for nonlocality, where probabilities do not conform to local expectations. Starting with the idea that nonlocality may be veiled as one of the ways cosmic censorship operates, we further argue that in order to preserve the ordinary objective reality described by local realism and general relativity, veiled nonlocality and cosmic censorship are indispensable operational aspects of the interactions of observers with physical systems. Nevertheless, it may be possible to indirectly detect traces of nonlocality in experiments where humans and other sentient subjects are involved.

## I. INTRODUCTION

The standard neuroscience view is to consider consciousness as an emergent property, which mediates awareness and remembrance [1]. In contrast, the subjective view, which is closely linked to the role of observation in quantum measurements, and specifically the Copenhagen Interpretation and its revision by von Neumann [2, 3, 4], is that consciousness provides the individual observer with agency and freedom. In von Neumann's view, Nature also exhibits free choice of response to an act of observation. The primary role of conscious observations *and* the preservation of an apparent local realism are implied by the lack of nonlocalities [5].

The realist position as encapsulated in the Many Worlds Interpretation (MWI) takes the wave function to be the underlying reality and so if nonlocalities existed, that would support the MWI view and global entanglements would prevail at all levels [6]. The epistemological and ontological consequences of such views are both counterintuitive and contrary to everyday observations of an objective reality. The apparent paradox of *freedom* of making observational choices *and* the corresponding freedom of response of Nature to specific observations [4] must be squared with unfoldment of Nature by physical law and this is the reason that consciousness is a controversial subject.

In the subjective view, most behavior is instinctive or driven by scripts, while the individual is free at creative moments. This dichotomous behavior creates its own paradoxes but it is consistent with the view that there are limits to what can be obtained by the scientific method and it is also consistent (as we will see) with the idea of universal complementarity. This should not be surprising since it is well known that logic, which is the basic tool that is used in the scientific methods, has its own paradoxes [7].

---

[a] kafatos@chapman.edu

The instinctive, automatic, and rational behavior is what is captured by the Dutch Book Arguments (DBA) [8] according to which an agent's beliefs should satisfy the axioms of probability. However, it is generally accepted that a creative act flies against rational or expected behavior, and as such, DBA would not apply in these circumstances. From a different perspective of the unity of consciousness held by Schrödinger and Wigner [9, 10], an agent cannot be considered to be a single self and, therefore, several belief systems will be associated with a specific behavior. In terms of behavior associated with the brain, one can speak of associational, reorganizational, and quantum languages since one can postulate physical elements with corresponding properties [11]. In the case of the brain, it would follow that the overall behavior of the biological system should therefore have aspects that integrate these languages in different way. For example, the nonlocal aspect of consciousness would be tied to the quantum language applicable in the physical brain.

The position against consciousness as an epiphenomenon was made forcefully by William James from the perspective of psychology [12]. He viewed consciousness as an active agent to be considered on the same footing as other organs of the body. We now know that "the conscious act of willfully altering the mode by which experiential information is processed itself changes, in systematic ways, the cerebral mechanisms used" [13]. To understand how effort, feeling, and will influence biological function requires going away from mechanistic conceptions of consciousness as an epiphenomenon [14].

New ideas from physics seem to pave the way for the consideration of consciousness as a fundamental category. This is so because quantum theory itself requires the consideration of the observer who interacts with the system causing it to change in a non-unitary manner [15]. Are there limits to the manner in which such observations are made that are essential to our understanding of the workings of consciousness?

In this paper we consider the relationship between veiled nonlocality [5] and cosmic censorship [16] principles. Since cosmic censorship deals primarily with gravitational singularities and veiled nonlocality deals with all kinds of nonlocal behavior, we argue that veiled locality is the more restrictive principle and it, therefore, subsumes the other. We further argue that the study of veiled nonlocality presents a fundamental way to bring in the non-unitary subjective perspective in the study of reality.

## II. QUANTUM THEORY AND CONSCIOUSNESS

The general applicability of the framework of complementarity beyond the confines of quantum world has been proposed in several works, e.g. Kafatos and Nadeau [17], Theise and Kafatos [18] for biology, and Roy and Kafatos [19] in brain dynamics. Consciousness also exhibits complementary and paradoxical aspects. For example, in several experiments the decision related to random choice is prefigured in electrical activity *before* the choice appears to have been made [20].

Given the parallels of subjective sense of consciousness and quantum theory [21] as well as the underlying principle of complementarity in both, it has been proposed that perhaps quantum theory will unlock the mystery of consciousness. This idea is supported by the quantum Zeno effect that shows how observation can guide the dynamics of a physical system [22]. The quantum Zeno effect is different than making an observational choice in standard von Neumann measurement theory. The quantum Zeno effect makes it possible for the observer to steer the state to one of his choice or to freeze it. For example, if a two-level system is observed, in the limit of continuous observations, the object is frozen forever in its initial state and it never evolves!

Observations through the quantum Zeno effect yield stable outcomes, which could be interpreted in some way that the observer becomes entangled with a particular quantum state through the act of observation, which is reminiscent of quantum nonlocality present in entangled pairs of photons or particles. As consciousness appears to work in a nonlocal manner, one might speculate that this nonlocality is similar to that of quantum theory.

A good setting to seek nonlocality in quantum theory is to study remotely situated entangled particles. That useful information cannot be sent remotely using such particles is called the no-signaling theorem [23]. In the experiments to investigate nonlocality, imperfections in the experiments leave various loopholes permitting explanations based on local realistic theories [24]-[26]. One of the present authors has proposed that the loopholes will never be closed and experimental verification of nonlocality that excludes local realistic explanations will not be found because of the *principle of veiled nonlocality* [5]. This principle parallels the cosmic censorship hypothesis [16] but it appears to be more sweeping in scope for the reason that cosmic censorship was conceived of in relation to singularities in gravitation whereas nonlocality is fundamental to the workings of quantum mechanics, which is generally believed to be the most fundamental of the theories of physics.

Locality and nonlocality are complementary aspects of physics and they characterize experimental setups and resulting worldviews of classical physics and quantum physics, respectively. As a complementary pair they seem to characterize an intuitive understanding of consciousness. As such, these two aspects may indeed be fundamental characteristics of the nature of reality. It is possible that this dichotomous pair play a role in complexity as in [18], tying observations to choices made by observers. Since complexity implies non-computability, we have a more natural and novel way to look at non-computability as tied to veiled nonlocality. If that is the case, then one can assert that observational selection is inherently and irreducibly coupled to observed systems (e.g. biological structures), tying quantum measurement processes with measurement processes in general, both driven by choices of conscious observers.

### III. COSMIC CENSORSHIP AND VEILED NONLOCALITY

As background, we consider the related weak and strong cosmic censorship hypotheses that assert that gravitational singularities (such as black holes) arising in general relativity cannot be experimentally observed. The weak cosmic censorship hypothesis was conceived by Roger Penrose in 1969 and posits that no naked singularities, other than the Big Bang singularity, exist in the universe.

According to the strong cosmic censorship hypothesis, the fate of all observers should be predictable from the initial data and boundary conditions. The two conjectures are mathematically independent, as there exist spacetimes for which one of the two cosmic censorship hypotheses is valid and the other is violated.

The idea of weak cosmic censorship has two complementary aspects:

- Scientific theories must conform to the nature of the mind which has both local and nonlocal aspects. The local aspects give rise to the classical view of the cosmos and classical neural networks of the brain account for such local aspects of the mind. On the other hand, if the brain is subject to quantum processes, then quantum nonlocality would also be applicable and the

- ability of the mind to perceive nonlocality, not bound by space and time, would be tied to quantum nonlocality.

- For classical general relativity, the preclusion of naked singularities conforms to locality as observed by distant observers. Matter falling onto a black hole can be followed all the way until crossing of the event horizon, but the existence of this horizon does not allow any observation of the central singularity.

The complementary aspect would be naked singularity that is implied (as is nonlocality) but cannot be observed in classical observations as it is covered by the black hole event horizon.

Penrose [16] hints that the issue of existence of naked singularities will not be completely resolved until we have a theory of quantum gravity, as the existence of black hole evaporation could possibly lead to observations of such singularities in semi-classical approaches to quantum gravity but would presumably not be there in a fully developed theory of quantum gravity.

Here, we assume a different approach than current efforts to unify gravity with quantum theory: Rather than trying to construct *ab initio* a common mathematical theory with all its known mathematical complexities, we instead choose to understand the common theoretical elements of both theories, expressed in the two principles that apply, veiled nonlocality and cosmic censorship, and thus gain the ability to proceed to unification. This would be similar to the historical development of general relativity starting from the Mach principle and the Principle of Equivalence.

The inability to directly observe nonlocality and naked singularities must be what gives the observed world a classical appearance, even though in the von Neumann picture, the entire world is quantum. At the limit of space-time leading to singular values, we have the implied gravitational singularity, which forms the limit of classical general relativity. At the opposite end, the limit of infinitely extended space-time, we have the implied nonlocality, the limit of classical interactions. As Kafatos and Nadeau [17] argued, all localities, whether linking space-like regions through EPR experiments to test Bell's inequality [27] as performed by several teams, particularly the teams of Aspect et al. [28] or the delayed choice experiments of Wheeler tying together events in different time regions [17] are all really special cases of what they term "Type III nonlocality" tying space-time together but which cannot be observed through observations of objects that reside in space-time.

Here we provide a *gedanken* experiment that provides insights into the close relationship of veiled nonlocality and cosmic censorship, as pertaining to bringing closer to each other quantum theory and general relativity: If nonlocality were not veiled, then in principle a naked singularity could be subjected to experimental observational tests.

One could imagine carrying out an EPR experiment wherein one of the entangled particles would be sent onto a black hole (Figure 1). The measurement of a quantum parameter (such as the polarization for a two photon setup) in a region far from the black hole would be tied to the polarization of the second photon approaching the singularity inside the black hole, thus revealing the existence of the naked singularity. As a maximally entangled pair, the two photons as emitted by the source are described by:

$$|\varphi_0\rangle = \frac{1}{\sqrt{2}}(|00\rangle + |11\rangle) \qquad (1)$$

where in expression (1), the joint states |00> and |11> are the tensor products |0> |0> and |1> |1>, respectively, and in an assumed convention here, |0> is "up" polarization, while |1> is "down" polarization.

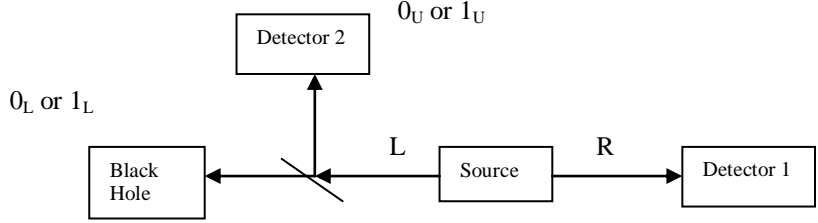

Figure 1. Linking veiled nonlocality and cosmic censorship

The left photon is put in a superposition over two paths by the half-silvered mirror serving as a beam splitter. The right (R) photon has its polarization measured along a particular direction. The left (L) photon gets into a superposition of the straight left and the up beams which will be represented by the Hadamard unitary transformation, or in terms of the following mappings:

$$|0_L\rangle \to \frac{1}{\sqrt{2}}(|0_L\rangle + |1_U\rangle) \qquad (2)$$

$$|1_L\rangle \to \frac{1}{\sqrt{2}}(|1_L\rangle - |0_U\rangle) \qquad (3)$$

The straight left photon moves onto a black hole and falls onto the singularity.

The photons after they have been put into a path superposition subsequent to the half-silvered mirror will be described by:

$$|\varphi_1\rangle = \frac{1}{2}(|0_L 0\rangle + |1_U 0\rangle + |1_L 1\rangle - |0_U 1\rangle) \qquad (4)$$

Now let us consider measurements at Detectors 1 and 2. As entangled particles, the von Neumann entropy of the system is zero and the entropy of each of the photons taken separately is one unit. Detector 1 will obtain the measurement of either 0 or 1, whereas there is probability 0.5 that Detector 2 measures a photon.

How to consider the situation when there is no measurement in Detector 2 which corresponds to the case that the photon was absorbed in the black hole singularity? There are two ways at looking at the situation:

- The information in the photon that went into the singularity is preserved somehow on the surface of the black hole as assumed in the holographic principle [29]. This allows for the nonlocal effect of the observation in Detector 1 on the photon in the singularity to be maintained but in a veiled manner. This approach is motivated by a desire to deal with the quantum information paradox but this approach is highly speculative.

- The information in the photon that has entered the singularity is actually destroyed. Therefore, overall unitarity is not maintained signifying that this represents a breakdown of quantum theory regime. If the black hole is represented by naked singularity then information in the left-straight path of figure 1 will be lost and this will correspondingly lead to non-unitary behavior in the left-upper path. This would be seen as nonlocal behavior that, in principle, can be determined but if it is not computationally possible to do so then this will also remain veiled.

The requirement of the appearance of a classical world constrains the observations of naked singularities and nonlocalities. As such, any aspects of reality that are not local would need to be put into that form by assuming additional hypotheses, in order to avoid global entanglement that covers everything in the universe and naked singularities. We note that direct observations of naked singularities and nonlocality will make it impossible to construct scientific theories that yield the appearance of an independently existing objective reality. We emphasize the term *direct* because of course both nonlocality and singularities are *implied* by quantum theory and general relativity, respectively.

Several scholars have argued that the only resolution to the information paradox problem associated with black holes is by the creation of new physics [29]. In a recent paper, Hui and Yang [30] show that by taking back-reaction into account, the paradox can be resolved without invoking any new physics beyond complementarity. We agree with this conclusion that no new physics is required, only the conjecture that the two principles are fundamentally tied together.

## IV. THE ROLE OF OBSERVATION AND TRACES OF NONLOCALITY AND CONSCIOUSNESS

The usual method of seeking evidence for nonlocality is whether Bell-type inequalities [31],[32] are violated. One would need to develop experimental arrangements with human (animal) subjects that can see evidence for this violation while excluding classical explanations for strong correlations that exist at distant locations.

The following chain of argument provides a broad review of the problem of consciousness and considers possible opportunities for advancing the science of consciousness:

1. Nonlocality is a correlate of the operation of thoughts in the mind;
2. Nonlocality is veiled;
3. Nonlocality may be observed indirectly.

In a certain sense the idea that both cosmic censorship and veiled nonlocality are necessary for the classical view of reality is a consequence of either the anthropic principle or any consideration in which the classical world comes together with conscious, living beings observing it. In the view of consciousness being the fundamental underlying reality, these principles are naturally necessary to provide for an external, objective reality.

If these principles did not exist, a very strange, globally-entangled, singularities-rich reality would exist, beyond the limits of any epistemology and ontology that we can imagine and even violating everyday experience. No observations of distinct objects, no scientific theories and, ultimately, no distinct conscious observers, can be conceived in this situation.

If indeed the consideration of these two principles is assumed to be tied to consciousness, it can provide the starting point for constructing an abstract theory dealing with consciousness.

## V. NONLOCALITY ASSOCIATED WITH A BLACK HOLE

Consider that an experimenter takes a pair of particles with entangled spins $|\uparrow\downarrow\rangle + |\downarrow\uparrow\rangle$ as in the consideration of Hawking radiation [33]. Each of these particles is a mixed state. Now if the experimenter injects the first particle into a black hole, the spin outside will continue to remain entangled with the one inside. Since the spin inside is trapped, when the black hole decays, the mixed state on the outside survives by itself which is problematic. Since the infalling particle loses its identity in the singularity and subsequently in the Hawking radiation emanating from the black hole, the outgoing particle is entangled both with the infalling particle as well as the Hawking radiation. This goes against the idea that entanglement cannot be shared with more than one entity. One way to get around it is to accept that the entanglement gets broken as the particle enters the black hole.

The principle of black hole complementarity [34] requires that different observers see the same bit of information differently. The external observer sees the horizon as a hot membrane that radiates information, while the infalling observer sees nothing. Black hole complementarity implies abandonment of locality and the equivalence principle which many find too much of a cost to be paid for the benefit it may provide.

Another resolution supposes that there is no entanglement between the emitted particle and Hawking radiation. But this implies violation of unitarity and black hole information loss. Most recently, Hawking [35] has abandoned the idea of event horizon in black holes. He proposes that gravitational collapse only produces apparent horizons. This proposal is consistent with the idea of AdS-CFT correspondence, that is correspondence of anti-de Sitter spaces (AdS) used in theories of quantum gravity with conformal field theories (CFT) that are quantum field theories. He claims that "the collapse to form a black hole will in general be chaotic and the dual CFT on the boundary of AdS will be turbulent. Thus, like weather forecasting on Earth, information will effectively be lost, although there would be no loss of unitarity."

Hawking's conjecture would lead to a different statement than it was stated in our preferred second choice that we presented following expression (4): In the case that the black hole has an event horizon, as expected in classical general relativity, unitarity would be destroyed, as we stated in the above second choice. Also, information in the photon would be destroyed. If, however, there is no event horizon, unitarity would be preserved, although information would be lost. Our *gendanken* setup could in principle be used to distinguish the two possibilities for a black hole, i.e. with or without event horizon. This would illustrate the close relationship between veiled nonlocality and cosmic censorship, linking quantum theory and gravity, or CFT and AdS.

Although the principle of information conservation, as arising out of unitarity, is at the basis of much contemporary physics theory, this may not be correct. If that is the case a deeper theory, different from quantum theory of which it is an approximation, should emerge.

## VI. CONCLUSIONS AND FUTURE WORK

In the paper it has been argued that there is an intimate relationship between veiled nonlocality, cosmic censorship and consciousness, enabling the emergence of a classical world with the existence of distinct observers and distinct objects, through distinct observations. Were it not for this intimate relationship, veiled nonlocality and cosmic censorship would be independent of each other and, we would argue, unification of general relativity and quantum theory, a difficult if not impossible task. In pursuing the Theory of Everything (TOE) that seeks to unify gravity with quantum theory, we argue that consciousness may have to be put into the matrix. As such, efforts to produce unification of physics will be tied to the issue of the role of the observer.

The paper also shows how nonlocality can be explained away in observations of classical systems just as naked singularities are explained away in observations in classical general relativity. In such observations, both nonlocality and naked singularities are seen as special limiting cases that cannot be realized in local explanations of observations. Are the distant correlations in the universe [36] (to explain which inflation has been postulated) examples of nonlocality at the cosmic level [cf. 17]? Our approach has the potential of providing new insights into the problem.

In the ER=EPR conjecture entangled particles are connected by wormholes [37],[38]. But such wormholes have not been found, which is supportive of the pervasiveness of veiled nonlocality. Can more persuasive experiments involving conscious observers *directly* interacting with quantum systems as related to nonlocality be performed? Perception (neuroscience) experiments seeking traces of nonlocality would provide an important scientific advance in the study of consciousness.